\documentstyle [12pt]{report}
\def\ZzZ{{\hbox{\tenrm Z\kern-.31em{Z}}}}
\def\CcC{{\hbox{\tenrm C\kern-.45em{\vrule height.67em width0.08em depth-
.04em
\hskip.45em }}}}
\def\mapright#1{\smash{\mathop{\longrightarrow}\limits^{#1}}}
\def\mapbelow#1{\smash{\mathop{\longrightarrow}\limits_{#1}}}

\newcommand{\ep}{\epsilon}
\newcommand{\lab}{\label}

\newcommand{\bc}{\begin{center}}
\newcommand{\ec}{\end{center}}
\newcommand{\be}{\begin{equation}}
\newcommand{\ee}{\end{equation}}
\newcommand{\bea}{\begin{eqnarray}}
\newcommand{\eea}{\end{eqnarray}}
\newcommand{\bs}{\begin{subequations}}
\newcommand{\es}{\end{subequations}}
\newcommand{\beq}{\begin{eqalignno}}
\newcommand{\eeq}{\end{eqalignno}}
\def\bol#1{\mbox{\boldmath\tiny $#1$\normalsize\unboldmath}}
\def\vec#1{\mbox{\boldmath $#1$\unboldmath}}
\def\bol#1{{\bf #1}}
\def\vec#1{{\bf #1}}

\newcommand{\half}{\frac{1}{2}}
\newcommand{\qrt}{\frac{1}{4}}

\def\Journal#1#2#3#4{{#1} {\bf #2}, {#3} {(#4)}}

\def\AP{\em Ann. Phys.}

\def\PRD{{\em Phys. Rev.}  \bf D}

\def\om{\omega}
\def\Om{\Omega}

\def\lab{\label}
\begin{document}
%


\vspace{2.0cm}
\bc
\Huge{Double Universe}  

\vspace{1.2cm}

\large{ E. Alfinito$^{*}$, R. Manka$^{\dag}$ and G. Vitiello$^{*}$} \\
\small
{\it ${}^{*}$Dipartimento di Fisica, Universit\`a di Salerno, 84100 
Salerno, Italy\\
and INFN Gruppo Collegato di Salerno}\\
{\it ${}^{\dag}$Physics Department, University of Katowice, Katowice, 
Poland}
\vspace{1.5cm}

\ec
\normalsize
{\bf Abstract} 

We discuss the canonical quantization of non-unitary time evolution in 
inflating Universe. We consider gravitational wave modes in the FRW metrics 
in a de Sitter phase and show that the vacuum is a two-mode 
SU(1,1) squeezed state of thermo field dynamics, thus exhibiting 
the link between inflationary evolution and thermal properties. In 
particular we discuss the entropy and the free energy of the system. The 
state space splits into many unitarily inequivalent representations 
of the canonical commutation relations parametrized by time $t$ and
non-unitary time evolution is described as a trajectory in the space 
of the representations: the system evolves in time by running over 
unitarily inequivalent representations. The generator of time evolution 
is related to the entropy operator. A central ingredient in our 
discussion is the doubling of the degrees of freedom which turns out to 
be the bridge to the unified picture of non-unitary time evolution, 
squeezing and thermal properties in inflating metrics.

\normalsize

\newpage

\setcounter{chapter}{1}
\setcounter{equation}{0}
\section*{1 Introduction}

Inflationary Universe scenarios have gained a central position in the 
interest of cosmologists and, more generally, of physicists interested in 
quantum aspects of General Relativity. In fact, a phase of primordial 
inflation explains (or seems to explain) some open problems in General 
Relativity such as the flatness and horizon problems or the so-called 
problem of quantum fluctuations \cite{1, GU, 2}. It is therefore of crucial 
relevance the possibility of describing in quantum field theory (QFT) the 
non-unitary time evolution implied by inflationary models. The purpose of 
the present paper is in fact the study of the canonical quantization of 
inflating time evolution for gravitational wave modes and of their thermal 
properties.

Quantization is by itself a particularly difficult matter in curved
space-time due to the ambiguity in the definition of a privileged frame. 
The same concept of {\it vacuum} state is meaningless in the presence of a 
curved space-time background. The particle production in a gravitational 
field \cite{Park}, the Hawking radiation \cite{5} and the generation of 
gravitational waves from vacuum in an expanding Universe \cite{Grish1, 
Grish2, Grish3} are indeed strictly related with the problem of properly 
defining the particle number operator in curved
space-time.

Even more serious is the problem of quantizing the inflating time evolution 
in the expanding Universe, since to the problem of the vacuum definition in 
curved space-time adds up the problem of quantizing the non-unitary
time evolution dynamics. Our purpose in this paper is in fact to focus 
our study on such a last question: the quantization problem related with 
inflationary time evolution.
As wee will see, such a problem requires the use of the full richness of
QFT, namely of the whole infinite set of unitarily inequivalent
representations of the canonical commutation relations. As already
stressed in ref. \cite{MSV} in connection with the problem of the
quantization of the matter field in curved space-time, and as clearly
stated and showed by Wald \cite{WA1, WA2}, QFT in curved
space-time does indeed requires, to be mathematically well
formulated, the use of all possible unitarily inequivalent Hilbert
spaces.
 
Since the pioniering works by Grishchuk, and Grishchuk and Sidorov  
\cite{Grish1, Grish2, Grish3} it is known that relic graviton states are 
squeezed states of the same kind of squeezed coherent states well known 
in quantum optics
\cite{QuOp, Yuen}. Two-mode squeezed states for relic gravitons were then 
discussed in ref. \cite{Prok}, also in connection with the formalism of the 
entropy associated to linearized perturbations \cite{BraProk}. In this 
paper our discussion, which will be in fact limited to gravitational wave 
modes in inflating Universe, will show how the occurrence of the squeezing 
phenomenon, and in particular of two-mode squeezing, is deeply related with 
the quantization problem of non-unitary time evolution, indeed; thus 
revealing a feature not directly emerging from the existing literature on 
the subject. We note that the dissipative (inflating) term in the 
gravitational mode equation (see eq. (\ref{p12})) is generally incorporated 
into the frequency term by using the conformal time variable $\eta$
\cite{Grish1, Grish2, Grish3}; this is a very useful computational 
strategy, however our purpose in this paper is to illustrate the 
subtelties of the canonical quantization for non-unitary time evolution 
and therefore we must explicitely take care of the dissipative effects. 
Only in this way the full structure of the state space will be revealed. 
Moreover, we obtain as an immediate consequence of our result a full set 
of thermodynamical properties of inflating evolution, including the 
entropy result of \cite{Prok}. As a matter of fact, we recover in the 
present context, the connection with thermal field theory in the 
formalism set up by Takahashi and Umezawa, called Thermo Field Dynamics 
(TFD) \cite{TFD, Um1, Um}, which anyway underlies many works (even if not 
referred to) since the paper by Israel \cite{Israel} on TFD of black 
holes (see also \cite{Jo}).

We will reach our conclusions by resorting in a crucial way to results on 
quantum dissipation obtained in the study of the canonical quantization of 
the damped harmonic oscillator taken as a prototype of dissipative systems 
\cite{QD}. In such a study, the quantization method for damping oscillator 
proposed by Feshbach and Tikochinsky \cite{FT} in quantum  mechanics (QM), was 
shown to lead to a proper canonical quantization for dissipative systems 
provided the QFT framework (and not the QM framework, see below) is used. 
An essential ingredient in the canonical formalism for quantum dissipation 
is the doubling of the system degrees of freedom, as proposed in fact by 
Bateman \cite{Bat} and by Feshbach and Tikochinsky \cite{FT}.

On the other hand, inflationary metrics also implies time-dependent 
frequency for the gravitational wave modes, and this leads us to extend the 
canonical quantization method for non-unitary time evolution so to include 
the quantization formalism for parametric oscillator \cite{Perel}. As we 
will see, and as one should expect \cite{Knight} (see also \cite{Kim1} with 
relation to the generalized invariant method), this naturally brings us to 
squeezed states. The connection with squeezing is also foreseen in the 
light of the proof of equivalence of damped oscillator states with 
squeezed states \cite{CTRV}.
 
The emergence of thermodynamical properties and the TFD character of the 
vacuum state has been already pointed out in the studies of quantum 
dissipation \cite{QD}. And in this respect the doubling of the degrees of 
freedom, on which also TFD is built upon \cite{TFD}, again has revealed to 
be a crucially useful tool, rich of physical content.

One reason why in TFD thermal aspects of QFT may be naturally treated is 
that in such a formalism all the statistical properties arise without any 
superimposed condition and all the states (included the vacuum) have a 
thermal valence. This is obtained by doubling the system degrees of 
freedom. Such a doubling allows to take into account not only the system 
under consideration but also the thermal bath (or environment) in which it 
is embedded. From an operatorial and algebraic point of view the doubling 
of the degrees of freedom is deeply related with the $C^{*}$-algebraic 
structure of the theory \cite{Ojima} (in particular with the Gel'fand-
Naimark-Segal (GNS) construction \cite{Bratteli}); more recently it has 
been pointed out that the algebraic structure of TFD is fully included in 
the quantum deformation of the associated Hopf algebra,
the doubling being related to the coproduct and the quantum deformation 
parameter being related to the non-zero value of the temperature 
\cite{vit}.

In the doubling formalism, developed for the damping phenomena in QFT 
\cite{QD} (see also \cite{DFV, SWV, iorio}) and here adopted for the 
present discussion, the ground state (vacuum) and the corresponding Fock 
space are labelled by the time $t$: $|0(t)>$ and ${\cal H}_{t}$, 
respectively. In the infinite volume limit, for different times, $t \neq 
t'$, $|0(t)>$ turns out to be ortogonal to $|0(t')>$ and ${\cal H}_{t}$ 
unitarily inequivalent to ${\cal H}_{t'}$. In this way one obtains the 
parametrization by $t$ of infinitely many representations ${\cal H}_{t}$ 
of the canonical commutation relations. The non-unitary character of time 
evolution implied by dissipation finds its description in the
non-unitary equivalence among the ${\cal H}_{t}$ representations at 
different $t$'s: the system state space thus splits into many unitarily 
inequivalent representations. We call the collection of the ${\cal H}_{t}$ 
representations the "representation space" \{${\cal H}_{t}$\}. Time 
evolution is then described as a trajectory in the space of the 
representations: the system evolves in time by running over unitarily 
inequivalent representations. 
We also show that the generator responsible 
for non-unitary time evolution is related to the entropy operator. This 
fact should not be surprising (see \cite{DFV, QD}) since dissipation 
(inflation) involves irreversibility in time evolution ({\it the arrow of 
time}). As a matter of fact, the system states are recognized to be 
SU(1,1) coherent states of the same kind of the thermal states of TFD, 
thus recovering a full set of thermal properties. The emerging picture of 
the canonical quantization scheme we obtain is thus a unified view of many 
features of inflationary evolution, including coherence, two-mode 
squeezing, entropy and vacuum thermal properties. It explicitely
shows how the algebraic construction utilizing the 
infinitely many unitarily
inequivalent representations works; in some sense it is the GNS
construction of $C^*$ algebra ``at work'' \cite{WA1, WA2, Bratteli}.

We perform the analysis of inflating metrics by using the quasi-linear 
approximation. In the de Donder gauge condition, Einstein equations lead, 
in the inflating case of Friedmann-Robertson-Walker (FRW) metrics, to the 
damped harmonic oscillator equation for the partial waves of the field $ 
h_{\mu\nu}$ \cite{Grish1, Grish2, Grish3} (for this reason, even when in our 
discussion we use the word dissipation we actually always refer to the 
inflating scenario). We then proceed with the quantization method for the 
damped oscillator mentioned above. The doubling of the degrees of freedom 
reflects itself on the metric structure in such a way we have the 
doubling of the $ h_{\mu\nu}$ partial waves. In this sense we speak of 
"double Universe".

The physical interpretation is that the doubled degrees of freedom 
introduce the {\it complement} to the inflating system, thus {\it 
closing} it, as required by the canonical quantization procedure. This 
points to the root of the mathematical difficulty in the canonical 
quantization of inflating evolution: we show in fact that non-unitary 
time evolution cannot be properly quantized if a single representation of 
the canonical commutation relations is available (as it happens in 
QM). However, the von Neumann theorem of QM does not 
hold in QFT, due to the infinite number of degrees of freedom, and the 
existence of infinitely many unitarily inequivalent representations of 
the canonical commutation relations is allowed. Then, the quantization of 
inflating systems may be performed by taking advantage of the full space of 
the representations, and this is achieved by closing the system (by 
doubling the degrees of freedom) and by allowing evolution by tunneling 
{\it through} the representations. As we said above, the emerging vacuum 
structure is the one of the squeezed coherent states. Squeezing has thus a 
deep dynamical origin and is related with the intrinsic group theory 
properties of the inflating evolution. The condensate vacuum structure is 
such that the difference between the  number of relic graviton modes and
of their doubled modes  
is constant in time (creation and annihilation occurs always in {\it 
pairs}) and the doubled mode can be interpreted as the {\it hole} for 
graviton mode. This reminds us of the similar situation occurring in the 
matter field particle creation in curved space-time background \cite{MSV}.

It is also interesting to observe that the damping term in the wave 
mode evolution equation is actually a manifestation of the curvature 
represented by the time-dependent inflating metrics; and that 
complementing the system in order to proceed to canonical quantization is 
actually equivalent to {\it shielding} the curvature effects due to such an 
inflating time-dependence. This is related with the {\it gauge} structure 
of quantum dissipation \cite{BGPV} and of TFD \cite{vit, Nakamura, Henning}.

The paper is organized as follows. In Sec. 2 we introduce general, 
well known features of the quasi-linear approximation leading in the 
inflating case to the damped parametric oscillator equation for the $ 
h_{\mu\nu}$ wave modes. In Sec. 3 we start the discussion of the system 
quantization, also showing there how the doubling of the degrees of 
freedom works in order to complement the system. In Sec. 4 we obtain the 
Hamiltonian spectrum by the method of the spectrum generating algebra and 
exhibit the theory vacuum structure, its time evolution and its two-mode 
squeezing character. We also show that the QM framework is not adeguate 
in view of the vacuum instability. In Sec. 5 we cure this pathology by 
moving to the QFT framework and we discuss the evolution as trajectories 
{\it over} unitarily inequivalent representations. Entropy, free energy and 
the first principle of thermodynamics are discussed in Sec. 6. It is 
intersting to remark that entropy appears in our formalism as the 
generator of the non-unitary time evolution and at the same time as the 
free energy responce to temperature variations; moreover, heat dissipation 
is related to variations in the condensate structure of the vacuum.

In this paper, since our interest is mainly focused on the possibility of 
setting up a canonical quantization scheme for inflating evolution 
exhibiting at once general features as squeezing and thermal properties, we 
do not consider more specific model features, or renormalization problems, 
neither we study symmetry restoration mechanisms due to thermal effects.
Concluding remarks are reported in Sec. 7.

\setcounter{chapter}{2}
\setcounter{equation}{0}
\section*{2 Inflating Universe}

In the four dimensional space-time $x^{\mu} = \{x_{0}= ct, x^{i}\},\, 
i=1,2,3, $ we consider the
so-called linear approximation where one decomposes the metrics 
$g_{\mu\nu}$ as \be g_{\mu\nu}\, =\,g^{0}_{\mu\nu}\,+\,h_{\mu\nu}.  
\lab{p1}\ee
When one chooses the flat background metrics
\be 
g_{\mu \nu }^0=\eta _{\mu \nu }=\left(  
\begin{tabular}{llll} 
1 & 0 & 0 & 0 \\  
0 & -1 & 0 & 0 \\  
0 & 0 & -1 & 0 \\  
0 & 0 & 0 & -1 
\end{tabular} 
\right)  ~~,
\lab{p2}\ee
as customary one defines
\be
\overline{h}_{\mu \nu }=h_{\mu \nu }-\frac 12\eta _{\mu \nu }h  
~~, \lab{p3}\ee
with $h\,=\, h^{\mu}_{\mu}$, which transforms as 
\be
\overline{h}_{\mu \nu }\rightarrow \overline{h}_{\mu \nu }^{\prime }=%
\overline{h}_{\mu \nu }-\partial _\mu \varsigma _\nu -\partial _\nu 
\varsigma _\mu +\eta _{\mu \nu }\partial _\lambda \varsigma ^\lambda  
\lab{p4}\ee
when 
\be
x^{\mu}\rightarrow x^{\prime \mu}=x^{\mu}+\varsigma ^{\mu}(x).  
\lab{p5}\ee
In the same way as in Q.E.D., the (de Donder) gauge condition
\be
\partial _\mu \overline{h}^{\mu \nu }=0  
\lab{p6}\ee
is satisfied provided the class of gauge functions $\varsigma ^\lambda $ 
solutions of 
\be
\Box \varsigma ^\lambda =0  \lab{p7}\ee
is adopted. In the gauge (\ref{p6}) the Einstein equations
\be
G_{\mu \nu }=R_{\mu \nu }-\frac 12g_{\mu \nu }R=0  
\lab{p8}\ee
give
\be
\Box \overline{h}^{\mu \nu }=0.  
\lab{p9}\ee
The field $\overline{h}^{\mu \nu }$ is then decomposed in partial waves
\be
\overline{h}^{\mu \nu }=\sum_\lambda \frac 1{(2\pi )^{\frac 32}}\int 
d^3 {\vec k}e_{{\vec k}\lambda }^{\mu \nu }\{u_{{\bol k}\lambda }(t)
e^{i{\bol k}\cdot  
{\bol x}}+u_{{\bol k}\lambda }^{\dag}(t)e^{-i{\bol k}\cdot{\bol x}
}\}   
\lab{p10}\ee
with 
$k\equiv(k_0=\omega =ck,{\vec k})$ . The wave function
$u_{{\bol k}\lambda }(t)$  
satisfies the 
simple harmonic oscillator equation
\be
\frac{d^2}{dt^2}u_{{\bol k}\lambda}(t)+\omega^{2} 
u_{{\bol k}\lambda}(t)=0  ~~. 
\lab{p10a}
\ee

When, instead of $\eta_{\mu\nu}$, the
metrics
\be
g_{\mu \nu }^0=\left(  
\begin{tabular}{llll} 
1 & 0 & 0 & 0 \\  
0 & -$a^2$(t) & 0 & 0 \\  
0 & 0 & -$a^2$(t) & 0 \\  
0 & 0 & 0 & -$a^2$(t) 
\end{tabular} 
\right)  
\lab{p11}\ee
with
\be
a(t)=a_0e^{\frac 13H t} ,
\lab{p15}\ee
and $H \,=\, const. = \frac{3\stackrel{.}{a}(t)}{a(t)} $ (the Hubble 
constant),
is adopted, $\overline{h}^{\mu \nu }$ may be still decomposed in partial waves
as in (\ref{p10}); 
however, in such a case one obtains the equation for the parametric 
(frequency time-dependent) damped harmonic oscillator \cite{1, Grish1, 
Grish2, Grish3} (see also \cite{GG})
\be
\stackrel{..}{u}(t)+H \stackrel{.}{u}(t)+\omega ^2(t)u(t)=0  
\lab{p12}\ee
with  
\be
{{\omega}^2}(t)=\frac{k_3^{2}c^2}{a^2(t)}~~.
\lab{p14}\ee
In Eq.(\ref{p12}) we have used $u(t)\,\equiv\,u_{{\bol k}\lambda}(t)$. In the 
Minkowsky
space-time $\om$ is constant in time, but when the Universe expands, 
$\om$ is time-dependent $\om = \om(t)$. This can be seen as a generalization 
of the Doppler effect.

The dissipative (inflating) term $H \stackrel{.}{u}$ 
in eq. (\ref{p12}) is generally incorporated into the frequency term by 
using the conformal time variable $\eta$
\cite{Grish1, Grish2, Grish3}; as observed in the introduction such a
computational strategy is very 
useful in the phenomenological approach, however our purpose in this paper 
is to illustrate the subtelties of the canonical quantization for non-unitary 
time evolution and therefore we must explicitely take care of the 
dissipative term in eq. (\ref{p12}). Only in this way the full structure of 
the state space will be revealed.

It is interesting to remark that eq.(\ref{p12}) with $a(t)$,  $H$ and 
$\om(t)$ given by (\ref{p15}) and (\ref{p14})
is also known as Hill-type equation \cite{Hill}. By setting $\xi = \ep \, 
z$, $\ep = \frac{3k_{3}c}{a_{0} H}$, $z= e^{-\frac{H}{3}t}$ and $u= z^{2} V$
we can easily show that
it can be written in the form of the spherical Bessel 
equation 
\be
\xi^{2} V_{\xi\xi} + 2 \xi V_{\xi} + \left(\xi^{2}-2\right)V = 0
~~. \lab{p17}\ee

One particular solution of eq.(\ref{p17}) is indeed the spherical Bessel
function of the first kind
\be
j_{1}(\xi)= \frac{\sin \xi}{\xi^{2}} - \frac{\cos \xi}{\xi} 
~~, \lab{p18}\ee
i.e.
\be
u(t)=\frac{1}{\ep^2} \left[\sin(\ep e^{-\frac{H}{3}t}) 
-  \ep e^{-\frac{H}{3}t}\, \cos(\ep e^{-\frac{H}{3}t})\right] 
~~,\lab{p19}\ee
which in fact is solution of (\ref{p12}).

\setcounter{chapter}{3}
\setcounter{equation}{0}
\section*{3 Quantization of inflating time evolution}

In ref.\cite{QD} it has been discussed the quantization of the 
one dimensional damped harmonic oscillator with constant frequency, i.e. 
of eq. (\ref{p12}) with ${\omega}^{2}(t) = const.$, and it has been shown 
that the canonical quantization can be properly performed by doubling the 
degrees of freedom of the system and by working in the QFT framework. 
The physical reason to double the degrees of 
freedom of the dissipative (damped) system relies in the fact that one must 
work with closed systems as required by the canonical quantization 
formalism. On the other hand, since the system states split into unitarily 
inequivalent representations of the canonical commutation relations 
\cite{QD}, one is forced to use QFT where infinitely unitarily inequivalent 
representations indeed exist (QM is not adeguate due to the Von Neumann 
theorem that states that all the representations are unitarily equivalent 
for systems with finite number of degres of freedom; see the discussion in 
\cite{QD}). Therefore, in order to perform the canonical quantization of 
the oscillator (\ref{p12}), according to \cite{QD} we consider the double 
oscillator system
\be\stackrel{..}{u}+H \stackrel{.}{u}+\omega ^2(t)u=  0 ~,\lab{p20}\ee 
\be \stackrel{..}{v}-H \stackrel{.}{v}+\omega ^2(t)v=  0 ~,
\lab{p21}
\ee
and in this sense we speak of a "double Universe". We observe indeed that
in the same way eq. (\ref{p20}) is implied by the inflating metrics, eq.
(\ref{p21}) for the oscillator $v$ can be associated to the "deflating" 
metrics
\be
{\tilde g}_{\mu \nu }^0 =\left(
\begin{tabular}{llll}
1 & 0                     & 0 & 0 \\
0 & $-$$\frac 1{a^2(t)}$  & 0 & 0 \\
0 & 0 & $-$$\frac 1{a^2(t)}$  & 0 \\
0 & 0 & 0 & $-$$\frac 1{a^2(t)}$
\end{tabular} \right)
\lab{p21b}\ee
with $a(t)$ given by (\ref{p15}).  We thus 
may think of the oscillator $v$ as associated to the "deflating" Universe, 
complementary to the inflating one. Note that ${\tilde g}_{\mu \nu }^0 = 
(g_{\mu \nu }^0)^{-1}$ and that the D'Alembert operator leading to 
(\ref{p21}) is given by $- \sqrt{-g}{\partial }^{\mu}{1\over{\sqrt{-g}}} 
g_{\mu \nu}{\partial }^{\nu} $ with $g \equiv det(g_{\mu \nu})$.

In order to proceed in our discussion it is now convenient to write down 
the Lagrangian in terms of $u$ and $v$ modes from which eqs.(\ref{p20}), 
(\ref{p21}) are directly obtained:
\be L=\stackrel{.}{u}\stackrel{.}{v}+\frac 12H (u\stackrel{.}{v}-
v\stackrel{.}{u})-\omega ^2(t)uv  ~.
\lab{p30}\ee
The canonical momenta are
\be p_u=\frac{\partial 
L}{\partial \stackrel{.}{u}}=\stackrel{.}{v}-\frac 12
H v ~,
\lab{p31}\ee
\be p_v=\frac{\partial L}{\partial 
\stackrel{.}{v}}=\stackrel{.}{u}+\frac 12H u ~,
\lab{p32}\ee
and the Hamiltonian is
\be {\cal H} = p_{u} \dot u + p_{v} \dot v - L =  p_{u} p_{v} + 
{{1}\over{2}}
H\left ( v p_{v} - u p_{u} \right ) + \left ( {\omega}^{2}(t) -{{H 
^{2}}\over{4}} \right ) u v \quad .  \lab{p33}\ee
We will put $\Omega(t) \equiv \left [ \left ({\omega}^{2}(t)  - {{H 
^{2}}\over{4}} \right ) \right ]^{1\over{2}}$, which we will get real, 
i.e. ${\omega}^{2}(t) > {{H ^{2}}\over{4}}$~: i.e. $0 \leq t < 
{3\over{H}}ln({2k_{3}c\over{a_{0}H}})$ , with
${2k_{3}c\over{a_{0}H}} \geq 1$; 
the same reality condition is however also fulfilled for $k_{3} \geq
{a_{0}H\over{2c}}e^{{H\over{3}}t}$ which tells us that as time flows
the reality condition excludes long wave modes, which in turn acts as an
intrinsic infrared cut-off, very welcome, as we will see, for the
well-definiteness of the operator fields. We therefore will adopt such a
last constraint in our subsequent discussion. See also \cite{1} 
for a discussion of time domains in inflating models. 

It is interesting to observe that $v = u e^{Ht}$ is solution of (\ref{p21})
and that by setting $u(t) = {1\over {\sqrt 2}} r(t)e^{{-Ht\over 2}}$ and 
$v(t) = {1\over {\sqrt 2}} r(t)e^{{Ht\over2}}$ the system of equations
(\ref{p20}) and (\ref{p21}) is equivalent to the equation
\be
\stackrel{..}{r} +\Omega ^2(t)r=  0 ~,
\lab{p21c}
\ee
which is the equation for the parametric oscillator $r(t)$ (see also 
\cite{BGPV}). This further clarifies the meaning of the doubling of the $u$ 
oscillator: the $u-v$ system is a non-inflating (and non-deflating) system. 
This is why it is now possible to set up the canonical quantization scheme.

We introduce indeed as usual the commutators
\be[\, u , 
p_{u} \, ] = i\, \hbar = [\, v , p_{v} \,] \quad , \quad  [\, u , v \,] = 0 
= [\, p_{u} , p_{v} \, ] ~~, \lab{p34}\ee
and it turns out to be convenient to introduce the variables $U$ and $V$ by 
the transformations
\be U(t)\, =\, \frac{u(t) + v(t)}{\sqrt 2}, \qquad 
V(t)\, =\, \frac{u(t) - v(t)}{\sqrt 2} 
~~, \lab{p35}\ee
which preserve the commutation relations (\ref{p34}):
\be[\, U , p_{U} \, ] = i\, \hbar = [\, V , p_{V} \,] \quad , \quad  [\, U , 
V \,] = 0 = [\, p_{U} , p_{V} \, ] ~.
\lab{p36}\ee

In terms of the variables $U$ and $V$ it now appears that we are dealing 
with the decomposition of the parametric oscillator $r(t)$ on the {\it 
hyperbolic} plane (i.e. in the pseudoeuclidean metrics): $r^{2}(t) = 
U^{2}(t) -V^{2}(t)$.
 
The Lagrangian (\ref{p30}) is rewritten in terms of $U$ and $V$ as
\be
L= L_{0,U} - L_{0,V}  + {H \over 2}({\dot U} V - {\dot V} U) ~~, \lab{p37}\ee
with
\be L_{0,U} =  {1\over 2} \dot U^2 - {{\omega^2}(t)\over 2} U^2 ~~,
~~~L_{0,V} =  {1\over 2} \dot V^2 - {{\omega^2}(t)\over 2} V^2 ~~.
\lab{p38}\ee
The associate momenta are
\be
p_U =  \dot U + {H \over 2} V ~~,\qquad
p_V = - \dot V -{H \over 2} U ~~, \lab{p39}\ee
and the motion equations corresponding to the set (\ref{p20})-
(\ref{p21}) are
\be
 \ddot U + H \dot V + {\omega^2}(t) U  = 0  ~~,\lab{p40-a}\ee
\be
 \ddot V + H \dot U + {\omega^2}(t) V  = 0  ~~. \lab{p40-b}\ee

The Hamiltonian (\ref{p33}) becomes
\be
{\cal H} = {\cal H}_U - {\cal H}_V = {1 \over 2} (p_U - {H \over 2}V)^2
+ {{\omega^2 (t)}\over 2} U^2 - {1 \over 2} (p_V + {H \over 2}U)^2 -
{{\omega^2 (t)}\over 2} V^2 ~~. \lab{p41}\ee

Eq.(\ref{p37}) shows that the inflation (or dissipative) 
 $H$-term actually acts as a coupling between the oscillators $U$ 
and $V$ and produces a correction to the kinetic energy for both 
oscillators (cf.eq.(\ref{p41})).
In the following we will see that the group structure underlying the
Hamiltonian (\ref{p41}) is the one of SU(1,1).
The occurrence of
the minus sign between $L_{0,U}$ and $L_{0,V}$ and
$H_{U}$ and $H_{V}$ , respectively, is in fact directly related with the 
SU(1,1) group structure.

As already observed above, our problem also involves the quantization of 
the parametric oscillator. We then proceed to the quantization of 
(\ref{p41}) by resorting to the quantization method of the parametric 
harmonic oscillator described in ref. \cite{Perel} and to the quantization 
method for the damped oscillator of ref.\cite{QD}. To this aim it is better 
to write the Hamiltonian (\ref{p41}) in the form:
\be {\cal H} = {1 \over 2} {p_U}^2 + {1 \over 2}{\Omega}^2(t)U^2
-{1 \over 2} {p_V}^2 - {1 \over 2}{\Omega}^2(t)V^2- {H \over 2}
({p_U} V + {p_V} U) ~.
 ~~ \lab{p42}\ee
We introduce the annihilation and creation 
operators:
\be
A= \frac{1}{\sqrt{2}} \left(\frac{p_U}{\sqrt{\hbar\om_{0}}}-iU
\sqrt{\frac{\om_{0}}{\hbar}}
\right), \qquad\qquad
A^{\dag}= \frac{1}{\sqrt{2}} \left(\frac{p_U}{\sqrt{\hbar\om_{0}}} + iU
\sqrt{\frac{\om_{0}}{\hbar}}
\right), \lab{p43}\ee
\be
B= \frac{1}{\sqrt{2}} \left(\frac{p_V}{\sqrt{\hbar\om_{0}}}-
 iV\sqrt{\frac{\om_{0}}{\hbar}}
\right), \qquad\qquad
B^{\dag} = \frac{1}{\sqrt{2}} \left(\frac{p_V}{\sqrt{\hbar\om_{0}}} + 
 iV\sqrt{\frac{\om_{0}}{\hbar}}
\right),
\lab{p44}\ee
with commutation relations
\be
[A,A^{\dag}] = 1 = [B,B^{\dag}], \quad [A,B] = 0, \quad 
[A^{\dag},B^{\dag}] = 0 ~,
\lab{p45}\ee
and all other commutators equal to zero. In eqs. (\ref{p43}), (\ref{p44})
${\omega}_{0}$ denotes an arbitrary frequency introduced according to the 
quantization procedure for parametric oscillator 
discussed in ref. \cite{Perel}. By putting $x = U, V$, we indeed have 
(cf. eq.(\ref{p42})):
\be
{1 \over 2} {p_x}^2 + {1 \over 2}{\Omega}^2(t)x^2 =
{\Om}_{0} K_{0}^{x} - {\Om}_{1} K_{1}^{x}
\lab{p45c}\ee
with
\be
\Omega_{0,1}= \om_{0} \left(\frac{\Om^{2}(t)}{\om^{2}_{0}}\pm 1\right) ~~,
\lab{p46}\ee
and
\be
{K_{0,1}}^x = \frac{1}{2 \om_{0}}\left(\frac{p_{x}^{2}}{2} \pm 
\frac{x^{2}}{2}
\om^{2}_{0}
\right) ~,
\lab{p48}\ee
which together with
\be
{K_{2}}^x = \frac{1}{4}(p_{x}x+xp_{x})
\lab{p481}\ee
close the $su(1,1)$ algebra:
\be
[K_{1}^x,K_{2}^x]=-i K_{0}^x, \quad [K_{2}^x,K_{0}^x]=i K_{1}^x, \quad
[K_{0}^x,K_{1}^x]= i K_{2}^x ~.
\lab{p53}\ee

Of course,
$[K_{i}^x ,K_{j}^{x'}] = 0$ for any $i,j$ and $x \neq x'$.
By using eqs. (\ref{p43})-(\ref{p44}) and eqs. (\ref{p48}) and 
(\ref{p481}) we introduce then the operators
$$
K_{0} = \half (A^{\dag}A -B^{\dag} B ) 
\qquad \qquad
K_{1 }= \qrt \left[\left(A ^{2}
+ {A^{\dag} }^{2} \right)  - \left(B ^{2}
 + {B^{\dag} }^{2}  \right)\right] , $$
\be K_{2 }= i\qrt \left[ \left(A ^{2}
 - {A^{\dag} }^{2}  \right) + \left( B ^{2}
 -{B^{\dag} }^{2}  \right)\right] , \lab{p55}\ee
which also close the algebra $su(1,1)$:
\be
[K_{1},K_{2}]=-i K_{0}, \quad [K_{2},K_{0}]=i K_{1}, \quad
[K_{0},K_{1}]= i K_{2},
\lab{p53b}\ee
We further remark that
\be
J_{+ }= A^{\dag} B^{\dag} , \quad  J_{- }= 
A B , \quad
J_{0 }= \frac{1}{2}(A^{\dag} A +
B^{\dag} B  + 1),
\lab{p56}\ee
with $J_{1 } = \half (J_{+ } + J_{- })$ and
$J_{2 } = -{i \over 2} (J_{+ } - J_{- })$ ,
also close the $su(1,1)$ algebra. Note that
\be
2iJ_{2 } =
  A^{\dag} B^{\dag}  - A B
\lab{p561}\ee
commutes with each $K_{i} ,~i=0,1,2$ (eqs. (\ref{p55})), and that
\be
{\cal C}  \equiv \frac{1}{2}(A^{\dag} A  -B^{\dag}B ) = {K_{0 }}
~~ \lab{p57}\ee
commutes with each $J_{i} ,~i=0,+,-$ (eqs. (\ref{p56})). $2iJ_{2}$ and 
$\cal C$ are indeed (related to) the Casimir operators for the algebras of 
generators (\ref{p55}) and (\ref{p56}), respectively.

In terms of $A $ and
$B $ the Hamiltonian (\ref{p42}) is finally written as
\be
{\cal H}  = {\cal H}_{0} + {\cal H}_{I_1} + {\cal H}_{I_2}
\lab{p58}\ee
\be
{\cal H}_{0} = \half \hbar \Om_{0 }(t)
(A^{\dag}  A  - B^{\dag}  B  )
= \hbar \Om_{0 }(t){\cal C}  =
\hbar \Om_{0 }(t) K_{0 } ~,
\lab{p59}\ee
\be
{\cal  H}_{I_1} = - {1\over 4}\hbar \Om_{1 }(t)
\left[\left(A ^{2}
+ {A^{\dag} }^{2} \right) - \left( B ^{2} + {B^{\dag}}^{2} \right)\right]
= - \hbar \Om_{1 }(t)K_{1 } ~,
\lab{p60}\ee
\be
{\cal H}_{I_2} = i  {\Gamma}
{\hbar} \left(A^{\dag}  B^{\dag}  -A B  
\right) = i \hbar {\Gamma }
( J_{+ } - J_{- }) ~.
\lab{p61}\ee
with ${\Gamma}  \equiv {{H}  \over 2}$.

We note that for any t
\be
[{\cal H}_{0}, {\cal H}_{I_2}] = 0 = [{\cal H}_{I_1} , {\cal H}_{I_2}]~,
\lab{p62}\ee
which guaranties that under time evolution the minus sign appearing in 
${\cal H}_{0}$ is not
harmful (i.e., once one starts with a positive definite Hamiltonian it 
remains
lower bounded).

Finally, we recall that the $A$ and $B$ operators (and all other operators) 
as well as other quantities, e.g. ~${\omega}(t)$, actually are dependent on 
the momentum $\vec k$ (and on other degrees of freedom) and thus our 
formulas should be understood as carrying such a $\vec k$ labels which we 
have been omitting for simplicity.
For instance the commutators (\ref{p45}) are indeed to be understood 
as
\be
[A_{\bol k},A^{\dag}_{{\bol k}'}] = {\delta}_{{\bol k},{\bol k}'} = 
[B_{\bol k},B^{\dag}_{{\bol k}'}], \quad [A_{\bol k},B_{{\bol k}'}] = 0, 
\quad [A^{\dag}_{\bol k},B^{\dag}_{{\bol k}'}] = 0 ~.
\lab{p51}\ee

We also observe that the operators ${K_{i,{\bol k}}}$,
$i = 0,1,2$ (or $i= 0,+,-$), for fixed ${\vec k}$ close the algebra 
$su_{\bol k}(1,1)$ and that they commute for any $i,j$ for ${\vec k} \neq
{\vec k}'$.

Our next task is to study the Hilbert space structure and this will be done 
in the following section.

\setcounter{chapter}{4}
\setcounter{equation}{0}
\section*{4 The vacuum structure}

Again, for simplicity, we will omit the $\vec k$ indeces whenever no 
misunderstanding arises. They will be restored
at the end. 

In order to study the eigenstates of the Hamiltonian (\ref{p58}), we 
consider the set  $\{ |n_{A} , n_{B} > \}$  of simultaneous
eigenvectors of $A^{\dagger} A$ and $B^{\dagger} B$, with $n_{A}$, $n_{B}$
non-negative integers.
These are eigenstates of ${\cal H}_{0}$ with eigenvalues
${1 \over 2}\hbar \Om_{0}(t)
(n_{A} -n_{B})$ for any $t$.  The eigenstates of ${\cal H}_{I_2}$ can be 
written in the standard
basis, in terms of the eigenstates of ${\left ( J_{3} -{1\over{2}} \right 
)}$ in the representation labelled by the value $j \in
Z_{1\over{2}}$ of ${\cal C}$,  $\{ | j , m > \, ; \, m \geq |j| \}$: \be
{\cal C} | j , m > = j | j , m > \quad , \quad j = {1\over{2}} (n_{A} - 
n_{B}) \quad ;
\ee
\be
 \left ( J_{3} - {1\over{2}} \right ) | j , m > = m | j , m > \quad , \quad 
m = {1\over{2}} (n_{A} + n_{B}) \quad .  \lab{p71}
\ee

By using the $su(1,1)$ algebra
\be
[\, J_{+} , J_{-}\, ] = - 2 J_{3} \quad ,
\quad [\, J_{3}  , J_{\pm}\, ] = \pm 
J_{\pm} \quad , \lab{p72}
\ee
one can show that
the kets $| \psi_{j , m} > \equiv
e^{\left ( + {{\pi}\over{2}} J_{1} \right )} | j , m >$
satisfy indeed the equation $J_{2} | \psi_{j , m} > = \mu | \psi_{j , m}>$ 
with pure imaginary $\mu \equiv i \left ( m + {1\over{2}} \right ) $.

Notice that although $J_{2}$ appears to be hermitian, it has
pure imaginary discrete spectrum in $| \psi_{j , m} >$. This apparent 
contraddiction is related with the
well known \cite{FT, agi} fact that the states $| \psi_{j , m} >$ are not 
normalizable and
the transformation generating $| \psi_{j , m} >$ from $|j , m >$ is a
non-unitary transformation in $SU (1,1)$ (it is not a proper rotation in 
$SU(1,1)$, but is rather
a pseudorotation in $SL ( 2 , C )$ \cite{QD}).  In other words
 $|\psi_{j , m} >$ does not
provide a unitary irreducible representation ({\sl UIR}), consistently  
with the fact \cite{ln} that in any {\sl UIR}
of $SU(1,1)$ $J_{2}$ should have purely continuous and real spectrum 
(which does not happen in the case of $| \psi_{j , m} >$). However,
{\sl UIR} with continuous and real spectrum are not adeguate for the 
description of dissipating and inflating phenomena and therefore 
they are of no help to us.

By following refs. \cite{QD,FT} we can bypass this difficulty by 
introducing in the Hilbert
space a new metric with a suitable inner product in such a 
way that
$| \psi_{j ,m} >$ has a finite norm. To this aim we consider the
antiunitary operation ${\cal T}$ under which
$(A , B) \mapright{\cal T} ( - A^{\dagger} , -B^{\dagger})$ and 
introduce the conjugation
$< \psi_{j , m} | \equiv$ $[\, {\cal T} | \psi_{j 
 , m} >\, ]^{\dagger}$, with
${\cal T} | \psi_{j , m} > \equiv$ $| \psi_{j , - (m+1)} >$.
The hermitian of
$J_{2} | \psi_{j , m} > = \mu | \psi_{j , m}> ~,~
\mu \equiv i \left ( m + 
{1\over{2}} \right )$, is now 
\be
< \psi_{j , - (m+1)} | J_{2} = \mu_{\cal T} < \psi_{j , - (m+1)} | \quad , 
\quad \mu_{\cal T} = - i \left [\, - (m+1) + {1\over{2}}\, \right ] = 
 - \mu^{*} = \mu\quad .
\lab{p73}\ee

We are left with the discussion of the eigenstates of  ${\cal 
H}_{I_{1}}$. We can "rotate away" \cite{so} ${\cal H}_{I_1}$
 by using the transformation ${\cal H}
\rightarrow {{\cal H}^{\prime}} \equiv e^{i\theta(t) K_{2}}{\cal H}
e^{-i\theta(t) K_{2}}$ with $\tanh {\theta(t)} = -{\Om_1 (t) \over \Om_0 
(t)}$ at any $t$.
We obtain
\be
{{\cal H}^{\prime}} \equiv e^{i\theta(t) K_{2}}{\cal H}
e^{-i\theta(t) K_{2}} = {\cal H}^{\prime}_{0} + {\cal H}_{I_{2}}
~~, ~~~ \tanh {\theta(t)} = -{\Om_1 (t) \over \Om_0 (t)}~,
\lab{p74}\ee
with
\be
{{\cal H}^{\prime}}_{0} = \hbar \Om(t)
(A^{\dag} A - B^{\dag} B )
\lab{p75}\ee
\be
[{{\cal H}^{\prime}}_{0} , {\cal H}_{I_{2}} ] = 0 ~.
\lab{p752}\ee
Here we have used the algebra (\ref{p53b}), eq. (\ref{p46}) and the fact 
that ${\cal H}_{I_2}$ commutes with
$K_{2}$. Note that when the $\vec k$ indeces are restored we
have ${\theta(t)} \equiv {\theta}_{\bol k}(t)$. Also note that the choice
$\tanh {\theta(t)} = -{\Om_1 (t) \over \Om_0 (t)}$ is allowed since the 
modulus of
$(-{\Om_1 (t) \over \Om_0 (t)})$ is at most equal to one (for any $\vec k$ 
and any $t$).

In conclusion, the eigenstates of the Hamiltonian $\cal H$ at $t$, eq.
(\ref{p58}), are states of type
$e^{-i\theta(t) K_{2}} | \psi_{j , m} >$~:
$$
{\cal H}e^{-i\theta(t) K_{2}} | \psi_{j , m} > =
e^{-i\theta(t) K_{2}} e^{i\theta(t) K_{2}}
{\cal H}e^{-i\theta(t) K_{2} }| \psi_{j , m} > =
$$
\be
e^{-i\theta(t) K_{2}}({\cal H}^{\prime}_{0} + {\cal H}_{I_2})
|\psi_{j , m} > = \left( \hbar \Om (t) (n_A - n_B) -  i \hbar {\Gamma}(n_A + 
n_B + 1) \right)
e^{-i\theta(t) K_{2} }| \psi_{j , m} > ~~,
\lab{p76}\ee
where we have been using the algebras (\ref{p53b}), (\ref{p72})
and the commuting properties of the related Casimir operators. We note 
that $\Omega (t)$ appears to be the common frequency of the two oscillators 
$A$ and $B$. In conclusion, the spectrum of ${\cal H}$ is determined by 
using the so-called method
of the spectrum generating algebras \cite{Perel, so}.

The solution to the Schr\"odinger equation can be given with 
reference to the initial time pure state $| j , m_{0} >$ (see refs. 
\cite{QD,FT}). When in particular, the initial state, say at arbitrary 
initial time $t_{0}$, is the {\it vacuum} for ${{\cal H}^{\prime}}_{0}$, 
{\it i.e.} $| n_{A} = 0 , n_{B} = 0 > \equiv |0>$, with $A |0> = 0 = 
B|0>$ ({\it i.e.} $j=0 , m_{0}=0$ for any $\vec k$),
the state 
\be
|0(\theta(t_{0}))> = e^{-i\theta (t_{0}) K_{2} }|0> ~,
\ee
at $t_{0}$ 
(and for given ${\vec k}$), is the zero energy eigenstate
(the {\it vacuum}) of ${\cal H}_{0} + {\cal H}_{I_1}$ at $t_{0}$:
\be
({\cal H}_{0} + {\cal H}_{I_1})|_{t_{0}}|0(\theta (t_{0}))> =
e^{-i\theta (t_{0}) K_{2} }{\cal H}^{\prime}_{0}|0> = 0 ~~ 
for~~any~~arbitrary~~t_{0}.
\lab{p77}\ee
Notice that ${\cal H}^{\prime}_{0}|0> = 0$ for any $t$, but that
$({\cal H}_{0} + {\cal H}_{I_1})|0> \neq 0$ and ${\cal H}^{\prime}_{0} 
|0(\theta (t_{0}))> \neq 0$ for any $t$ . However,
expectation values of ${\cal H}^{\prime}_{0}$ and of
$({\cal H}_{0} + {\cal H}_{I_1})$ in $|0>$ and in $|0(\theta (t_{0}))>$ are
all zero at any $t$ (see also below).

The state $|0(\theta (t))>$ is a generalized $su(1,1)$
coherent state, appearing, as well known, in the study of the parametric
exicitations of the quantum oscillator \cite{Perel}.

In the following for simplicity we will put $t_{0} = 0$ and set $ 
\theta(t_{0}=0) \equiv \theta$ and $|0(\theta (t_{0}=0))> \equiv
|0(\theta)> $.

Since $A$
and $B$ are commuting operators we can write
\be
\exp\left({-i\theta K_{2} }\right) = \exp{{{\theta} \over 4}
\left(A^{2} - {A^{\dag}}^{2}  \right)}
\exp{{{\theta} \over 4} \left( B^{2} -{B^{\dag}}^{2}  
\right)}~,
\label{paa}\ee
i.e. $\exp\left({-i\theta K_{2} }\right)$
can be factorized as the product of two (commuting) 
single-mode squeezing generators and for this reason 
we will
refer to the state $|0(\theta)>$ as to 
the squeezed
vacuum (at this level actually it is not, strictly speaking, a squeezed 
state since squeezed states are obtained by applying the squeezing 
generator to a (Glauber-type) coherent state).

Time evolution of the squeezed vacuum $|0(\theta)>$ is given by
\be
|0({\theta},t) > = \exp{\left ( - i t {{{\cal H}}\over{\hbar}}\right )} 
|0(\theta)> = \exp{\left ( - i t {{{\cal H}_{I_2}}\over{\hbar}}\right )}
|0(\theta)> ~,
\lab{p78}\ee
due to (\ref{p62}) and to (\ref{p77}). Notice that in eq.(\ref{p78}) 
${\cal H}$ denotes the Hamiltonian at time $t_{0}=0$, {\it i.e.} the 
Hamiltonian (\ref{p58})  with
$\Omega_{0}(t_{0}=0)$ and $\Omega_{1}(t_{0}=0)$.  Note that
(\ref{p78}) can be written also as
\be
|0({\theta},t) >  = \exp{\left ( -i\theta K_{2} \right ) } 
\exp{\left ( - i t {{{\cal H}_{I_2}}\over{\hbar}}\right )} |0>
=  \exp{\left( -i\theta K_{2} \right)} |0(t)>.\lab{p781}\ee

We can now better appreciate the advantage of explicitely
taking care of the  dissipative term in eq. (\ref{p12}): the
contributions from the non-unitary 
evolution term and from the frequency term are now separated and explicitely
shown in eqs. (\ref{p78}) and (\ref{p781}). 

Let us study $|0({\theta},t) >$ given by (\ref{p78}). We observe that the
operators $A$ and $B$ transform under
$\exp{\left ( -i\theta K_{2} \right )}$ (for any given ${\vec k}$) as
\be
A  \mapsto A (\theta) = {\it e}^{
-i\theta K_{2} }A  {\it e}^{i\theta K_{2}}
 =  A  \cosh{(\half\theta )} + A ^{\dagger} \sinh{(
\half\theta )}~,
\lab{p7261}\ee
\be
B  \mapsto B (\theta) = {\it e}^{
-i\theta K_{2} }B  {\it e}^{i\theta K_{2} }
 =  B  \cosh{(\half\theta )} + B ^{\dagger} \sinh{(
\half\theta )}~.
\lab{p7262}\ee
These transformations
are the well known squeezing 
transformations which preserve the commutation relations (\ref{p45}) (and 
(\ref{p51})).
One has
\be
A (\theta) |0(\theta)> = 0 = B (\theta) |0(\theta)> 
\quad, \quad
 \lab{p27}\ee
and the number of modes of type $A $ in the state $|0(\theta)>$
is given, 
by
\be
n_{A }(\theta) \,\equiv\, 
< 0(\theta) | A ^{\dagger} A  | 
0(\theta) > \,=\, \sinh^{2}\bigl( \theta  \bigr) \quad ;
\lab{p272}\ee
and similarly for the modes of type $B $.

We also observe that the commutativity of $J_2$ with $K_{2}$ ensures that
\be
 A^{\dag} B^{\dag}  - A B  =
  A^{\dag} ({\theta})B^{\dag} ({\theta}) - A({\theta})B ({\theta})~. 
 \lab{p28}\ee

By using (\ref{p74}), (\ref{p7261}) and (\ref{p7262}) we also obtain
$$
{\cal H}_{0} + {\cal H}_{I_{1}} + {\cal H}_{I_{2}} =
e^{-i\theta K_{2}}{{\cal H}^{\prime}}_{0}e^{i\theta K_{2}} +
{\cal H}_{I_{2}} =$$
\be
 = \hbar \Om(0)
(A^{\dag}(\theta) A(\theta) - B^{\dag}(\theta) B(\theta) ) + i\hbar \Gamma
( A^{\dag} ({\theta})B^{\dag} ({\theta}) - A({\theta})B ({\theta}))~, 
\lab{p281}\ee

We have (at finite volume $V$)
\be
|0(\theta,t)> = {1\over{\cosh{(\Gamma  t)}}} \exp{
\left ( \tanh {(\Gamma  t)} J_{+}(\theta) \right )} |0(\theta)> \quad , 
\lab{p79}\ee
with $J_{+}(\theta) \equiv A ^{\dagger}(\theta) B^{\dagger}(\theta)$, 
namely a $su(1,1)$  generalized coherent state (a two mode Glauber-type 
coherent
state) with equal numbers of modes $A (\theta)$ and
$B (\theta)$ 
condensed in it (for each $\vec k$) at each $t$. At time $t$ ~ 
$|0({\theta},t)>$
is thus a proper squeezed state.  We observe that
\be
<0(\theta,t) | 0(\theta,t)> = 1 \quad \forall t \quad , \lab{p710}\ee
\be
<0(\theta,t) | 0(\theta)> = \exp{\left ( - \ln \cosh {(\Gamma  t)} \right 
)} \quad ; \lab{p711}\ee
which shows how, provided ${\Gamma > 0}$ ,
\be
<0(\theta,t) | 0(\theta)> \, \propto \exp{\left ( - t  \Gamma \right )}
\rightarrow 0 \; ~for~large~t .
\lab{p712}\ee

Thus eq. (\ref{p712}) shows the vacuum instability: 
time evolution brings out of the initial-time Hilbert space for large $t$. 
This is not acceptable in quantum mechanics since there the Von Neumann 
theorem states that all the representations of the canonical commutation 
relations are unitarily equivalent and therefore there is no room in 
quantum mechanics for
non-unitary time evolution as the one in (\ref{p712}).
On the contrary, in QFT there exist infinitely many unitarily inequivalent 
representations and this suggests to us to consider 
our problem in the framework of QFT, which we will do in the next section.

\setcounter{chapter}{5}
\setcounter{equation}{0}
\section*{5 Quantum field theory of inflationary evolution}

To set up 
the formalism in QFT we have to consider the infinite volume limit; 
however, as customary, we will work at finite volume and at the end of 
the computations we take the limit $V \rightarrow \infty$. The QFT 
Hamiltonian is introduced as
\be
{\cal H}  = {\cal H}_{0} + {\cal H}_{I_1} + {\cal H}_{I_2}
\lab{ph1}\ee
\be
{\cal H}_{0} =  \sum_{\bol k} \half \hbar \Om_{0,{\bol k}}(t)
(A^{\dag}_{\bol k} A_{\bol k} - B^{\dag}_{\bol k} B_{\bol k} )
= \sum_{\bol k} \hbar \Om_{0,{\bol k}}(t){\cal C}_{\bol k} =
\sum_{\bol k} \hbar \Om_{0,{\bol k}}(t) K_{0,{\bol k}} ~,
\lab{ph2}\ee
\be
{\cal  H}_{I_1} = - \sum _{\bol k}{1\over 4}\hbar \Om_{1,{\bol k}}(t)
\left[\left(A_{\bol k}^{2}
+ {A^{\dag}_{\bol k}}^{2} \right) - \left( B_{\bol k}^{2} + {B^{\dag}_{\bol 
k}}^{2} \right)\right]
= - \sum_{\bol k} \hbar \Om_{1,{\bol k}}(t)K_{1,{\bol k}} ~,
\lab{ph3}\ee
\be
{\cal H}_{I_2} = i\sum _{\bol k} {\Gamma}_{\bol k}
{\hbar} \left(A^{\dag}_{\bol k} B^{\dag}_{\bol k} -A_{\bol k}B_{\bol k} 
\right) = i\sum_{\bol k} \hbar {\Gamma_{\bol k}}
( J_{+,{\bol k}} - J_{-,{\bol k}}) ~.
\lab{ph4}\ee
Notice that we have used $\vec k$-dependent $\Gamma$ and the 
relation between the ${\Gamma}_{\bol k}$'s and  ${\Gamma} 
\equiv {{H} \over 2}$ will be discussed below.

 We also have now
\be
{{\cal H}^{\prime}}_{0} = \sum_{\bol k}\hbar \Om_{\bol k}(t)
(A_{\bol k}^{\dag} A_{\bol k} - B_{\bol k}^{\dag} B_{\bol k} )~~.
\lab{ph5}\ee
At finite volume $V$, we formally have
\be
|0(\theta,t)> = \prod_{\bol k} {1\over{\cosh{(\Gamma_{\bol k} t)}}} \exp{
\left ( \tanh {(\Gamma_{\bol k} t)} J_{{\bol k}, +}(\theta) \right )} 
|0(\theta)> \quad , \lab{pq79}\ee
which corresponds to eq.(\ref{p79}). The state $|0({\theta},t)>$ is also a 
$su(1,1)$  generalized coherent state. Eqs. (\ref{p710})-(\ref{p712}) are 
now replaced by
\be
<0(\theta,t) | 0(\theta,t)> = 1 \quad \forall t \quad , \lab{pq710}\ee
\be
<0(\theta,t) | 0(\theta)> = \exp{\left ( - \sum_{\bol k} \ln \cosh 
{(\Gamma_{\bol k} t)} \right )} \quad ; \lab{pq711}\ee
which again exhibit non-unitary time evolution, provided ${\sum_{\bol k} 
\Gamma_{\bol k} > 0}$:
\be
<0(\theta,t) | 0(\theta)> \, \propto \exp{\left ( - t  \sum_{\bol k}  
\Gamma_{\bol k} \right )} \rightarrow  0 \; ~ for~large~t~ . \lab{pq712}\ee
Use of the customary
continuous limit relation $ \sum_{\bol k} 
\mapsto {V\over{(2 \pi)^{3}}} \int \! d^{3}{{\vec k}}$, 
for $\int \!
d^{3}{\vec k} \, \ln \cosh {(\Gamma_{\bol k} t)}$ 
finite and positive, gives in the
infinite volume limit
\be
<0(\theta,t) | 0(\theta)> \mapbelow{V \rightarrow \infty} 0 \quad \forall 
\, t \quad ,
\lab{p713}\ee
\be
<0(\theta,t) | 0(\theta',t') > \mapbelow{V \rightarrow \infty} 0 \quad 
with ~~ \theta' \equiv \theta (t_{0}'),~~\forall \, t\, , t'\, , t_{0}' 
\quad , \quad t \neq t'~~ .  \lab{p714}\ee

We conclude that in the infinite volume limit, vacua at $t$ and at $t'$,
~$\forall \, t ~ ,~ t' $ ~, with ~~ $t \neq t'$~, are orthogonal and the 
corresponding Hilbert spaces are unitary inequivalent representations
of the canonical commutation relations \cite{QD} (see also \cite{DFV}).

Under time evolution generated by ${\cal H}_{I_{2}}$ the operators $A_{\bol 
k}(\theta)$ and
$B_{\bol k}(\theta)$ transform as
\be
A_{\bol k}(\theta) \mapsto A_{\bol k}(\theta,t) = {\it e}^{- i 
{t\over{\hbar}} {\cal H}_{I_2}} A_{\bol k}(\theta) {\it e}^{i 
{t\over{\hbar}} {\cal H}_{I_2}} =  A_{\bol k}(\theta) \cosh{(\Gamma_{\bol k} 
t)} - B_{\bol k}^{\dagger}(\theta) \sinh{(
\Gamma_{\bol k} t)}~,
\lab{p7261a}\ee
\be B_{\bol k}(\theta) \mapsto B_{\bol k}(\theta,t) = {\it e}^{- i 
{t\over{\hbar}} {\cal H}_{I_2}} B_{\bol k}(\theta)
{\it e}^{i {t\over{\hbar}} 
{\cal H}_{I_2}} =  - A_{\bol k}^{\dagger}(\theta) \sinh{(\Gamma_{\bol k} t)} + 
B_{\bol k}(\theta) \cosh{(
\Gamma_{\bol k} t)} ~.
\lab{p7262a}\ee
These transformations are the Bogoliubov transformations and they can be 
understood as inner automorphism for the algebra $su_{\bol k} (1,1)$ and 
are canonical as they preserve
the commutation relations (\ref{p51}). Thus, at every $t$ we have a copy
$\{ A_{\bol k}(\theta,t),$ $A_{\bol k}^{\dagger}(\theta,t) ,
B_{\bol k}(\theta,t) ,
B_{\bol k}^{\dagger}(\theta,t) \, ; \, | 0(\theta,t) >\,
|\, \forall {\vec k} \}$ of the
original algebra and of its highest weight vector
$\{ A_{\bol k}(\theta), A_{\bol k}^{\dagger}(\theta) , B_{\bol k}(\theta) 
, B_{\bol k}^{\dagger}
(\theta) \, ; \, | 0(\theta) >\, |\,
\forall {\vec k} \}$, induced by the time evolution operator, i.e. we have a
realization of the operator algebra at each time t ( which can be 
implemented by Gel'fand-Naimark-Segal construction in the $C^{*}$-algebra 
formalism \cite{Bratteli}). The time evolution operator therefore acts as a 
generator
of the group of automorphisms of $\bigoplus_{\bol k}
su_{\bol k} (1,1)$ parametrized by time $t$.  We stress that
the copies of the original algebra provide unitarily inequivalent 
representations of the canonical commutation relations in the
infinite-volume limit, as
shown by eqs. (\ref{p714}).

At each time $t$ one has
\be
A_{\bol k}(\theta,t) |0(\theta,t)> = 0 = B_{\bol k}(\theta,t)
|0(\theta,t)> \quad,
\quad \forall 
\, t \quad ,
 \lab{p273}\ee
and the number of modes of type $A_{\bol k}(\theta)$ in the state 
$|0(\theta,t)>$ is given, at each instant $t$ by
\be
{n}_{A_{\bol k}}(t) \equiv < 0(\theta,t) | A_{\bol k}^{\dagger}
(\theta) 
A_{\bol k}(\theta) | 0(\theta,t) > =
< 0(t) | A_{\bol k}^{\dagger}A_{\bol k} | 0(t) >
= \sinh^{2}\bigl( \Gamma_{\bol k} t \bigr) \quad ;
\lab{p274}\ee
and similarly for the modes of type $B_{\bol k}(\theta)$. The state
$|0(t)>$ (see eq.(\ref{p781})) has an expression similar to (\ref{pq79}) 
with
$\theta$ missing.

We also observe that the commutativity of ${\cal C}$ (i.e. $K_{0}$) 
with ${\cal 
H}_{I_2}$ (i.e. $J_{2}$) 
ensures that the number $\left( n_{A_{\bol k}} - n_{B_{\bol k}} 
\right)\,$ is a constant of motion for any $\vec k$ and any $\theta$.
Moreover, one can show \cite{QD,TFD} that
the creation of a mode $A_{\bol k}(\theta,t)$ is equivalent to the 
destruction of a mode $B_{\bol k}(\theta,t)$ and vice-versa.  This means 
that the $B_{\bol k}(\theta,t)$ modes can be interpreted  as the {\sl 
holes} for the modes $A_{\bol k}(\theta,t)$: the $B$-system can be 
considered as the sink where the energy dissipated by the $A$-system flows.

Let us now observe that $|0(\theta)>$ is given by
\be
|0(\theta)> = \prod_{\bol k} {1\over{\cosh{(\theta_{\bol k})}}} \exp{
\left ( \tanh {(\theta_{\bol k})} j_{{\bol k}, +} \right )} |0> \quad , 
\lab{p792}\ee
with $j_{{\bol k} ,+} \equiv a_{\bol k}^{\dagger} b_{\bol 
k}^{\dagger} = {1\over2}({A_{\bol k}^{\dagger}}^{2} + {B_{\bol 
k}^{\dagger}}^{2})$, and $a_{\bol k} = {1\over{\sqrt 2}}(A_{\bol k} +
iB_{\bol k})$, $b_{\bol k} = {1\over{\sqrt 2}}(A_{\bol k} - iB_{\bol 
k})$. Thus, $|0(\theta)>$ also is a $su(1,1)$  generalized coherent state 
with equal numbers of modes $a_{\bol k}$ and
$b_{\bol k}$ condensed in it for each $\vec k$ and each $t$. We thus 
finally have
\be
|0({\theta},t)> = \prod_{{\bol k},{\bol q}} {1\over{\cosh{(\Gamma_{\bol k} 
t)}}{\cosh{(\theta_{\bol q})}} } \exp{
\left ( \tanh {(\Gamma_{\bol k} t)} J_{{\bol k}, +}(\theta) \right )} 
\exp{
\left ( \tanh {(\theta_{\bol q})} j_{{\bol q}, +} \right )} |0> \quad . 
\lab{p793}\ee

In a similar way we could derive the alternative expression for
$|0({\theta},t)>$:
\be
|0({\theta},t)> = \prod_{{\bol k},{\bol q}} {1\over{\cosh{(\Gamma_{\bol k} 
t)}}{\cosh{(\theta_{\bol q})}} } \exp{
\left ( \tanh {(\theta_{\bol q})} j_{{\bol q}, +}(t) \right )}
\exp{
\left ( \tanh {(\Gamma_{\bol k} t)} J_{{\bol k}, +} \right )} |0> \quad . 
 \lab{p794}\ee
where $J_{{\bol k}, +} = A_{\bol k}^{\dagger} B_{\bol 
k}^{\dagger}$, $j_{{\bol q}, +}(t) \equiv
a_{\bol k}^{\dagger}(t) b_{\bol k}^{\dagger}(t) = {1\over2}({A_{\bol 
k}^{\dagger}}^{2}(t) + {B_{\bol k}^{\dagger}}^{2}(t))$, and $a_{\bol 
k}(t) = {1\over{\sqrt 2}}(A_{\bol k}(t) + iB_{\bol k}(t))$, $b_{\bol 
k}(t) = {1\over{\sqrt 2}}(A_{\bol k}(t) - iB_{\bol k}(t))$, with 
commutators $[a_{\bol k}(t), a_{\bol q}^{\dagger}(t)] = {\delta}_{{\bol 
k},{\bol q}} = [b_{\bol k}(t), b_{\bol q}^{\dagger}(t)]$ and all other 
commutators equal to zero. The operators $A_{\bol k}(t)$ and $B_{\bol 
k}(t)$ are given by the (canonical) Bogoliubov transformations
\be
A_{\bol k} \mapsto A_{\bol k}(t) = {\it e}^{- i {t\over{\hbar}} {\cal 
H}_{I_2}} A_{\bol k} {\it e}^{i {t\over{\hbar}} {\cal H}_{I_2}} =  A_{\bol 
k} \cosh{(\Gamma_{\bol k} t)} - B_{\bol k}^{\dagger} \sinh{(
\Gamma_{\bol k} t)} \quad ,
\lab{795}\ee
\be
B_{\bol k} \mapsto B_{\bol k}(t) = {\it e}^{- i {t\over{\hbar}} {\cal 
H}_{I_2}} B_{\bol k} {\it e}^{i {t\over{\hbar}} {\cal H}_{I_2}} =  - 
A_{\bol k}^{\dagger} \sinh{(\Gamma_{\bol k} t)} + B_{\bol k} \cosh{(
\Gamma_{\bol k} t)} \quad .
\lab{p796}\ee

Finally, to establish the relation between
the ${\Gamma}_{\bol k}$'s and  ${\Gamma} 
\equiv {{H} \over 2}$ we note that in the continuum limit eqs. (\ref{p51})
become
\be
[A_{\bol k},A^{\dag}_{{\bol k}'}] = {\delta}({\bol k} - {\bol k}') = 
[B_{\bol k},B^{\dag}_{{\bol k}'}], \quad [A_{\bol k},B_{{\bol k}'}] = 0, 
\quad [A^{\dag}_{\bol k},B^{\dag}_{{\bol k}'}] = 0 ~,
\lab{p517}\ee
and that, as well known, the $A_{\bol k}$ (and $B_{\bol k}$) operators are 
not well defined on vectors in the Fock space; for instance $|A_{\bol k}> 
\equiv A_{\bol k}^{\dag}|0>$ is not a normalizable vector since from eqs. 
(\ref{p517}) one obtains $<A_{\bol k}|A_{\bol k}> = \delta ({\bol 0})$ 
which is infinity.
As customary one must then introduce wave-packet (smeared out) operators 
$A_{f}$ with spatial distribution described by square-integrable 
(ortonormal) functions
\be
f(\bol x) = {1 \over{(2\pi)^{3}}}\int {d^3{\bol k}} f({\bol k}) e^{i{\bol 
k}{\bol x}}
\lab{p798}\ee
i.e.
\be
A_{f} = {1 \over{(2\pi)^{3/2}}}\int {d^3{\bol k}} A_{\bol k} f({\bol k})
\lab{p797}\ee
with commutators
\be
[A_{f},A^{\dag}_{g}] = (f,g) = [B_{f},B^{\dag}_{g}],
 \quad [A_{f},B_{g}] = 0, \quad [A^{\dag}_{f},B^{\dag}_{g}] = 0 ~,
\lab{p518}\ee
with $(f,g)$ denoting the scalar product between $f$ and $g$. Now 
$<A_{f}|A_{f}> = 1$ and the $A_{f}$'s are well defined operators
in the Fock space in terms of which the observables have to be
expressed. In this connection it is interesting to recall that the
reality condition on $\Omega(t)$ (see sec. 3) naturally introduces the
infrared cut-off smearing out the operator fields. In conclusion, we
express the number operator as
\be
{n}_{A_{f}}(t) \equiv < 0(\theta,t) | A_{f}^{\dagger}(\theta) A_{f}(\theta) 
| 0(\theta,t) > =
 {1 \over{(2\pi)^{3}}}\int {d^3{\bol k}}
\sinh^{2} \bigl( \Gamma_{\bol k} t \bigr) |f({\bol k})|^{2}
\equiv \sinh^{2} \bigl( \Gamma t \bigr) ~ ,
\lab{p2746}\ee
and similarly for the modes of type $B_{f}(\theta)$ (cf. with 
eq. (\ref{p274})). Eq. (\ref{p2746}) specifies the relation between the 
${\Gamma}_{\bol k}$'s and ${\Gamma}$ and says that the number of $A_{f}$ 
modes does not depend on the volume.

The results obtained in this section clearly show the role of 
dissipation and its interplay with the time-dependent frequency term in
eq. (\ref{p12}). By using the conformal time coordinate as usually done
in the literature we would not be able to reveal the underlying rich
structure of the state space. 
Such a structure naturally leads us to recognize the thermal properties of
$|0({\theta},t)>$. This will done in the following section.

\setcounter{chapter}{6}
\setcounter{equation}{0}
\section*{6 Entropy and free energy in inflating Universe}

The vacuum state $|0({\theta},t)>$ as given by equation (\ref{pq79}) can be 
written as
\be
|0({\theta},t)>\, = \exp{\left ( - {1\over{2}} {\cal S}_{A({\theta})} 
\right )} |\,{\cal I}({\theta})>\, = \exp{\left ( - {1\over{2}} {\cal 
S}_{B({\theta})} \right )} |\,{\cal I}({\theta})> \quad ,
\lab{p81}\ee
where
\be
|\,{\cal I}({\theta})>\, \equiv \exp {\left( \sum_{\bol k}
A_{\bol k}^{\dagger}({\theta})
B_{\bol k}^{\dagger}({\theta}) \right)} |0({\theta})> =
\exp{\left ( -i\theta K_{2} \right ) }|{\cal I}> ~~,
\lab{p82}\ee
with $|{\cal I}>$ the invariant (not normalizable) vector ~\cite{TFD}
\be
|{\cal I}> \equiv \exp {\left( \sum_{\bol k}
A_{\bol k}^{\dagger}B_{\bol k}^{\dagger} \right)} |0> ~~,
\lab{p821}\ee
and
$$
{\cal S}_{A({\theta})} \equiv - \sum_{\bol k} \Bigl \{
A_{\bol k}^{\dagger}({\theta}) A_{\bol k}({\theta}) 
\ln \sinh^{2} \bigl ( \Gamma_{\bol k} t \bigr ) - A_{\bol k}({\theta}) 
A_{\bol k}^{\dagger}({\theta}) \ln \cosh^{2} \bigl ( \Gamma_{\bol k} t
\bigr ) \Bigr \} = $$
\be
=\, \exp{\left ( -i\theta K_{2} \right ) }{\cal S}_{A}\exp{\left
(i\theta K_{2} 
\right ) }~~.
\lab{p83}\ee
Here ${\cal S}_{A}$ is given by
\be
{\cal S}_{A} \equiv - \sum_{\bol k} \Bigl \{
A_{\bol k}^{\dagger} A_{\bol k} \ln \sinh^{2} \bigl ( \Gamma_{\bol k} t 
\bigr ) - A_{\bol k} A_{\bol k}^{\dagger} \ln \cosh^{2} \bigl ( 
\Gamma_{\bol k} t
\bigr ) \Bigr \}~~.
\lab{p831}\ee

${\cal S}_{B({\theta})}$ (${\cal S}_{B}$) is given by the same expression 
with $B_{\bol k}
({\theta})$ ($B_{\bol k}$) and $B_{\bol k}^{\dagger}({\theta})$
($B_{\bol k}^{\dagger}$) replacing $A_{\bol k}(\theta)$ ($A_{\bol k}$) and
$A_{\bol k}^{\dagger}({\theta})$ ($A_{\bol k}^{\dagger}$), 
respectively.  In the following we 
shall simply write ${\cal S}({\theta})$ (${\cal S}$) for either ${\cal 
S}_{A({\theta})}$
or ${\cal S}_{B({\theta})}$ (${\cal S}_{A}$ or ${\cal S}_{B}$).

 We also note that from eq. (\ref{p81}) we obtain 
\be
|0({\theta},t)> = \sum_{n \geq 0} \sqrt{W_{n}(t)}\, | n(\theta) , 
n(\theta) > \quad ,
\lab{p84}\ee
where $n(\theta)$ denotes the multi-index $\{ n_{\bol k}(\theta) \}$, and
\be
W_{n}(t) = \left ( \prod_{\bol k} {{\cosh^{2(n_{\bol k}+1)} \bigl ( 
\Gamma_{\bol k} t \bigr )}\over{\sinh^{2 n_{\bol k}} \bigl ( \Gamma_{\bol k} t 
\bigr )}} \right )^{-1} \quad , \quad 0 < W_{n} < 1 \quad .  
\lab{p85}\ee
Notice that the expansion (\ref{p84}) contains only terms for which 
$n_{A_{\bol k}({\theta})}$ equals $n_{B_{\bol k}({\theta})}$
for all $\vec k$'s and that
\be
\sum_{n \geq 0} W_{n}(t) = 1 \quad \quad {\rm for\; any}\; t,
\lab{p86}\ee
\be
<0(\theta,t)| {\cal S}({\theta}) |0(\theta,t)> =
<0(t)| {\cal S} |0(t)> =
- \sum_{n \geq 0} W_{n}(t) 
\ln{W_{n}(t)} \quad . \lab{p87}\ee
Eq. (\ref{p87}) leads us therefore to interpreting
${\cal S}({\theta})$ as the {\it 
entropy} for the dissipative system ~\cite{QD, TFD}.
We also observe that  $<0(\theta,t)| {\cal S}({\theta}) |0(\theta,t)>$
grows monotonically 
with $t$: the entropy for both $A$ and $B$
increases as the system evolves in time. Moreover, the difference ~~${\cal 
S}_{A(\theta)} - {\cal S}_{B(\theta)}$
is constant in time (cfr. (\ref{p281})):
\be
[\, {\cal S}_{A(\theta)} - {\cal S}_{B(\theta)} , {\cal H} 
] = 0 \quad
\lab{p88}\ee
(and, correspondingly, $[\,
{\cal S}_{A} - {\cal S}_{B} , {\cal H}' \,] = 0 $, cfr. (\ref{p74})) .
Since the $B$-particles are the holes for the $A$-particles, ~~
${\cal S}_{A(\theta)} -
{\cal S}_{B(\theta)}$~~ is in fact the (conserved) entropy for the closed
system.

Eqs. (\ref{p81}) and (\ref{p83}) show that the operational 
dependence of ${1\over{2}}
{\cal S}_{A({\theta})}$
(or respectively,  ${1\over{2}} {\cal S}_{B({\theta})}$) is uniquely on the 
$A$ ($B$) variables: thus in
eq. (\ref{p81}) time evolution is expressed solely in terms of the 
(sub)system $A$ ($B$)
with the elimination of the $B$ ($A$) variables. This reminds us of the
procedure by which one obtains the reduced density matrix by integrating
out bath variables.

For the time variation of $|0(\theta ,t)>$ at finite volume $V$, we obtain
\be
{{\partial}\over{\partial t}} |0(\theta ,t)> =  -  {1\over{2}} \left ( 
{{\partial {\cal S}({\theta})}\over{\partial t}} \right )
|0(\theta ,t)>  \quad .
\lab{p91}\ee

Equation (\ref{p91}) shows that $ {1\over{2}} \left ( 
{{\partial {\cal S}({\theta})}\over{\partial t}}  \right ) $ is the
generator of time-translations,
namely time evolution is controlled by the entropy variations. It is 
remarkable that the
operator that
controls time evolution is the variation of the dynamical variable
whose expectation value is formally an entropy:
these features reflect indeed correctly the
irreversibility of time evolution characteristic of dissipative (inflating) 
motion.
Dissipation (inflation) implies in fact the choice of a privileged
direction in time evolution ({\it time arrow}) with a consequent breaking of
time-reversal invariance.

We conclude that the system in its evolution runs over a variety of 
representations of the canonical commutation relations which are unitarily 
inequivalent to each other for $t \neq t'$ in the infinite-volume limit: 
the non-unitary character of time evolution implied by damping (inflation) 
is thus recovered,  in a consistent scheme, in the unitary inequivalence 
among representations at different times in the infinite-volume limit.

As already observed in the introduction, the similarities of the above 
analysis with results on the vacuum structure for {\it QFT} in curved
space-time and on the Hawking radiation for black-hole solutions 
~\cite{MSV}
suggest to us that the doubling of the degrees of freedom is intimately 
related with the non-trivial metric structure of space-time, the doubled 
degree of freedom signalling the lost of the Poincar\'e invariance.

We want now to further analyze the thermal concepts and properties of the 
formalism above presented.

We observe that the statistical nature of dissipative (inflating) phenomena 
naturally emerges from our formalism, even though no statistical concepts 
were introduced a priori (we have seen that the vacuum structure naturally 
leads to the {\it entropy operator} as time evolution generator (see 
eq.(\ref{p91})). We therefore ask ourselves whether such statistical 
features may actually be related to thermal concepts. We know from ref. 
~\cite{QD} (which here we closely follow) that this is indeed the case.
 
For the sake of
definiteness, let us consider the $A$-modes alone and introduce the 
functional
\be
F_{A} \equiv <0({\theta},t)| \Bigl ( {{\cal H}^{\prime} }_{0,A(\theta) } -
 {1\over{\beta}} {\cal S}_{A({\theta})} \Bigr ) |0({\theta},t)> =
<0(t)| \Bigl ( {{\cal H}^{\prime} }_{0,A} - {1\over{\beta}} {\cal 
S}_{A} \Bigr ) |0(t)> \quad .
\lab{p92}\ee
Here ${{\cal H}^{\prime}}_{0,A(\theta)} \equiv \sum_{\bol k} E_{\bol k} 
A_{\bol k}^{\dagger}(\theta) A_{\bol k}(\theta)$ and
$ {{\cal H}^{\prime} }_{0,A} \equiv \sum_{\bol k} E_{\bol k}
 A_{\bol k}^{\dagger}A_{\bol k}$; $\beta$ is a strictly 
positive function of time to be determined and $E_{\bol k} \equiv
\hbar \Omega_{\bol k}(t_{0}=0) - \mu$, with $\mu$ the chemical potential.

We write ${\sigma}_{\bol k} \equiv \Gamma_{\bol k} t$, and look for the 
values of $\sigma_{\bol k}$ rendering ${F}_{A(\theta)}$ stationary:
\be
{{\partial {F}_{A(\theta)}}\over{\partial \sigma_{\bol k}}} = 0 \quad ; \quad 
\forall \vec k \quad .
\lab{p93}\ee
\indent  Condition (\ref{p93}) is a stability condition to be satisfied for 
each representation. We now assume that $\beta$ is a slowly varying 
functions of t and thus eq. (\ref{p93}) gives
\be
\beta E_{\bol k} = - \ln \tanh^{2} \bigl ( \sigma_{\bol k} \bigr ) \quad .
\lab{p94}\ee
We have then
\be
{n}_{A_{\bol k}}(t) = \sinh^{2} \bigl ( \Gamma_{\bol k} t \bigr ) = 
{1\over{{\rm e}^{\beta (t) E_{\bol k}} - 1}} \quad , \lab{p101}
\ee
which is the Bose distribution for $A_{\bol k}$ at time $t$ provided
we assume $\beta (t)$ to represent the inverse temperature 
$\beta(t) = {1\over{k_{B} T(t)}}$ at time $t$ ($k_{B}$ denotes the Boltzmann
constant).  This allows us to recognize $\{ |0(\theta ,t)> \}$ as a 
representation of
the canonical commutation relations at finite temperature, equivalent
with the Thermo Field Dynamics representation $\{ |0(\beta )> \}$ of 
Takahashi and Umezawa ~\cite{TFD} . 

In conclusion we can interpret ${F}_{A}$ as the free energy and 
${n}_{A}$ as the average number of $A$-modes at the inverse temperature 
$\beta (t)$ at time $t$.

The change in time of the energy ${E_{A} \equiv \sum_{\bol k} E_{\bol k} 
{n}_{A_{\bol k}}}$ is given by
\be
d E_{A} ={{\partial}\over{\partial t}} \Bigl ( <0(t)| {\cal 
H}^{\prime}_{0,A} |0(t)> \Bigr ) d t  = \sum_{\bol k} E_{\bol k} 
\dot{n}_{A_{\bol k}}(t) d t \quad ; \lab{p981}\ee
and the change in the entropy by
\be
d S_{A} = {{\partial}\over{\partial t}} \Bigl ( <0(t)| {\cal S}_{A} |0(t)> 
\Bigr ) = \beta \sum_{\bol k} E_{\bol k}  
\dot{n}_{A_{\bol k}}(t) d t  = \beta d E_{A}(t) \quad ,
\lab{p982}\ee
provided we assume the changes in time of $\beta$ can be neglected
(which happens, {\it e.g.} in the case of adiabatic variations of 
temperature, at $T$ high enough) . Thus
we have
\be
d E_{A} - {1\over{\beta}} d {S}_{A} = 0 \quad ,
\lab{p983}\ee
consistently with the relation
obtained directly by minimizing the free energy
\be
d {F}_{A} = d E_{A} - {1\over{\beta}} d {S}_{A} \quad =0 ~~.
\lab{p984}\ee
Eq. (\ref{p984}) expresses the
first principle of thermodynamics for a system coupled with environment
at constant temperature and in absence of mechanical work and  it allows
us to recognize $E_{A}$ as the internal energy of the system.
The above discussion shows that time evolution induces transitions 
over inequivalent representations by inducing changes in the number of 
condensed modes in the vacuum. By defining as usual heat as
${dQ={1\over{\beta}} dS}$ we see that the change in time 
$\dot{n}_{A}$ of particles condensed in the vacuum turns out into heat 
dissipation $dQ$ \cite{QD}. Finally, (\ref{p984}) also shows that, provided 
variations of $E_{A}$ in the temperature are negligeable, entropy is as 
usual the free energy responce to temperature variations.

\setcounter{chapter}{6}
\setcounter{equation}{0}
\section*{7 Conclusions}

In this paper we have studied the canonical quantization of non-unitary 
time evolution in inflating Universe. We have shown that the vacuum is a 
two-mode squeezed state and we have discussed its thermal properties. We 
have considered only the gravitational wave modes in the FRW metrics in a 
de Sitter phase. We have shown that the vacuum turns out to be the 
generalized SU(1,1) coherent state of thermo field dynamics, thus 
exhibiting the link between inflationary evolution and thermal properties. 
In particular we have discussed the entropy and the free energy of the 
system recovering results similar to the ones presented in \cite{Prok}
(see also \cite{Kim2}).

A central ingredient in our discussion has been the doubling of the 
degrees of freedom, which we have imported from the canonical 
quantization procedure of the damped harmonic oscillator \cite{QD}. 
The doubling of the degrees of freedom is also a central tool in the TFD 
formalism of finite temperature QFT, and is thus the bridge to the 
unified picture of non-unitary time evolution, squeezing and thermal 
properties in inflating metrics.

{}From the thermal properties perspective the physical interpretation of the 
doubled degrees of freedom is the one of the thermal bath degrees of 
freedom; from the point of view of the vacuum structure the one of {\it 
holes} of the relic gravitons; form the hamiltonian formalism point of 
view the one of the {\it complement} to the dissipating (inflating) system.
In ref. \cite{SWV} the doubling formalism is shown to be related with the 
Feynman-Vernon \cite{FV} and with the Schwinger formalism \cite{Schw}.

We note that in ref. \cite{Prok}, even if not explicitely stated, the 
doubling of the degrees of freedom is actually introduced by considering 
the modes of momentum $\bf k$ and $- \bf k$ as the {\it couple} of modes of 
total zero momentum in terms of which the two-mode squeezed vacuum is 
described: the distinction between the
$\bf k$ and $- \bf k$ modes introduces a partition in the $\bf k$ space 
and leaves out the zero momentum modes which, 
although not entering in the condensate structures (in \cite{Prok} the 
summations are always limited to ${\bf k} > 0$), nevertless are present in 
the quantized field $\phi$ and in its canonical momentum $\pi$.
Also in \cite{Grish1, Grish2, Grish3} the doubling is actually present. In fact, 
in considering the general solution of the parametric oscillator $u(t)$
(we adopt here the same notation of ref. \cite{Grish3}) the authors consider 
{\it two} different representations of $u(t)$ in terms of two distinct 
basis, $\xi(\eta)$ and $\chi(\eta)$, respectively, with $\eta$
playing the role of time coordinate. In this way the doubling of the 
degrees of freedom is actually introduced, even if not mentioned. As a 
matter of fact, the doubling of the degrees of freedom is intrinsic to 
the Bogoliubov transformations, so that one 
deals with a doubled system anytime one works with such 
transformations. For this reason all the "mixed modes" formalisms 
(since Parker's work \cite{Park} ) necessarily involve the algebraic 
structure of the doubling of the modes.

We have shown that the system state space splits into many unitarily 
inequivalent representations of the canonical commutation relations 
parametrized by time $t$, ${\cal H}_{t}$, and non-unitary time evolution is 
then described as a trajectory in the space of the representations: the 
system evolves in time by running over unitarily inequivalent 
representations. This means that the full set of possible unitarily 
inequivalent Hilbert spaces must be exploited, which provides further 
support to known results in QFT in curved space-time \cite{MSV, WA1, 
WA2}. The generator of time evolution is related to the entropy operator, 
which indeed reflects the irreversibility in time evolution ({\it the 
arrow of time}). At the same time, entropy appears as the responce of 
free energy to the temperature variation and thus the intrinsic thermal 
character of the inflating evolution is exhibited.
In this connection, it is interesting to remark that a similar dynamical 
structure appears in the canonical quantization of matter field in a 
curved background \cite{MSV}, where the parametrization by proper time of 
the space of the representations allows the definition of the vacuum 
state and of the number operator. The present paper discussion and the 
results of \cite{MSV} point to the possibility of extending the 
present canonical quantization scheme also to the case of matter 
field in the background of expanding Universe. It is then an interesting 
question to ask how to incorporate in such a scheme the results on the
scalar field in an expanding geometry of refs. \cite{Guth, Eboli, Allen},
the self-similar time-dependent scale transformation for the functional 
Schr\"odinger equation for a free scalar field \cite{Vaut} and the 
effective potential models for inflating Universe \cite{Boya}.

As a further remark we would like to point out that the "negative" kinetic 
term in our Lagrangian structure (cf. section 3) also appears (in the 
absence of dissipation) in two-dimensional gravity models in conjunction 
with the problem of the ambiguity of the vacuum definition \cite{Jack}. In 
such a situation the choice of the annihilation operators discriminates 
between two different alternatives: either annihilation operators are 
associated with positive frequencies and then negative norm states appears, 
either annihilation operators competing to the positive kinetic term are 
associated to positive frequencies and annihilation operators competing to 
the negative kinetic term are associated to negative frequencies. In this 
last case no negative norm states appear and the canonical structure 
is similar to the one of quantum dissipation \cite{QD, BGPV} and of the 
inflationary time evolution described in the present paper.

Finally, we note that the algebraic structure of the doubling formalism 
can be related with the quantum deformation of the Weyl-Heisenberg 
algebra and that, as already mentioned, the TFD algebra is included in the 
deformed Hopf algebra. We thus expect that the quantum deformation 
mechanism plays a non-trivial role in the quantization procedure in 
expanding geometry. In particular, we expect \cite{iorio} that the 
deformation $q$-parameter is related with the inflating constant (the 
Hubble constant).

This work has been partially supported by INFN and by EU Contract ERB CHRX 
CT940423.


\end{document}